\documentclass{PoS}

\title{Recent progress in PDF theory}

\ShortTitle{Recent progress in PDF theory}

\author{\speaker{Marco Bonvini}\thanks{This work is supported by the Marie Sk\l{}odowska Curie grant HiPPiE@LHC.}\\
        INFN, Sezione di Roma 1, Piazzale Aldo Moro~5, 00185 Roma, Italy\\
        E-mail: \email{marco.bonvini@roma1.infn.it}}

\abstract{We discuss the recent results obtained in the context of PDF determination with the inclusion of logarithmic resummations and their phenomenological applications.}

\FullConference{Sixth Annual Conference on Large Hadron Collider Physics (LHCP2018)\\
                4-9 June 2018\\
                Bologna, Italy}

\usepackage{url}
\usepackage{cite}
\usepackage{blindtext}
\usepackage{booktabs}

\newcommand{\MSbar}{\overline{\rm MS}}
\newcommand{\as}{\alpha_s}
\newcommand{\Ord}{\mathcal{O}}

\begin{document}

\section{Logarithmic resummations in PDF theory}

Precise phenomenology at hadron colliders requires, among other things, precise and reliable parton distribution functions (PDFs).
On top of the inclusion of new and more precise data, refined fitting techniques, etc.,
one of the key ingredients to improve PDFs is the inclusion of higher order corrections in the theory used in their determination.
In addition to standard fixed-order corrections, resummations of some classes of logarithms play a special role in this task.
Indeed, on the one hand they offer a complementary way to go beyond the accuracy of fixed-order computations,
and on the other hand they are necessary in some kinematical regimes in order to stabilize the perturbative expansion,
which may otherwise be unreliable due the presence of large logarithms.

The resummation of threshold logarithms, important at large momentum fraction $x$,
has been explored in the context of PDF determination in Refs.~\cite{Corcella:2005us,Bonvini:2015ira}.
In the commonly adopted $\MSbar$ scheme, threshold resummation can affect partonic coefficient functions,
but not PDF (DGLAP) evolution.
Moreover, it is well known that the threshold logarithms in the coefficient functions make the perturbative expansion unstable
only at extremely large $x$, i.e.\ very close to threshold, while in a region of intermediate/large $x$ they
may be sizeable but behave in a perturbative way.
For this reason, for most of the large-$x$ data entering a PDF fit the inclusion of threshold resummation
has effectively the role of including approximate higher order corrections, thereby improving the theoretical description of these data.
Only a small amount of data points lie at very large $x$: for those points the resummation must be included, but the effect
on the resulting PDFs, though potentially sizeable, is invisible due to the large uncertainties in the $x\to1$ region.
As a result, the inclusion of threshold resummation in PDF determination has a modest impact at NLO,
because the (approximate) higher orders predicted by resummation are $\Ord(\as^2)$ and thus sizeable,
while the impact is much reduced at NNLO, where the higher order $\Ord(\as^3)$ correction induced by the resummation are smaller.

\begin{figure}[t]
  \centering
  \includegraphics[width=0.49\textwidth,page=1]{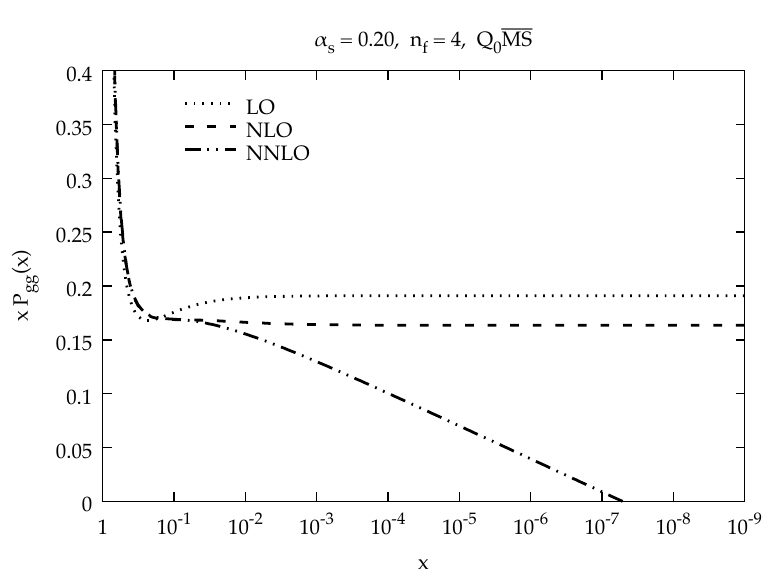}
  \includegraphics[width=0.49\textwidth,page=1]{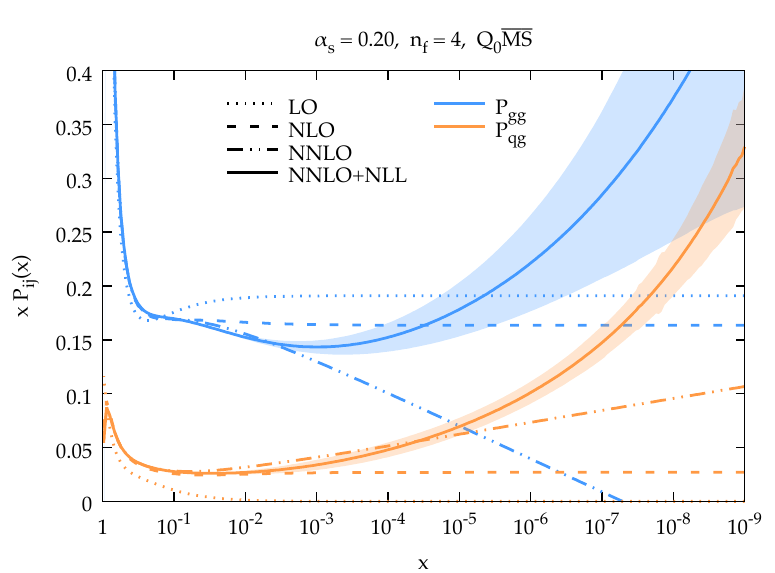}
  \caption{Left: fixed-order $P_{gg}$ splitting function at small-$x$.
    Right: $P_{gg}$ and $P_{qg}$ splitting functions including NLL resummation matched to NNLO.
    Results are for $\as=0.2$, $n_f=4$ in the $Q_0\MSbar$ scheme.}
  \label{fig:Pgg}
\end{figure}

The situation is drastically different in the opposite regime, namely at small $x$.
The instability of the perturbative expansions induced by large small-$x$ logarithms is much more severe,
and in $\MSbar$-like schemes it affects both coefficient functions and DGLAP splitting functions.
For instance, in Fig.~\ref{fig:Pgg} (left) we show the $P_{gg}$ splitting function at LO, NLO and NNLO,
for $\as=0.2$ corresponding to $Q\sim5$~GeV. We see that starting from NNLO, a logarithmic contribution
makes the perturbative expansion unreliable. Specifically, for $x\sim 10^{-3}$ (at this scale)
the NNLO correction is already larger than the NLO correction. Similar considerations also hold for DIS coefficient functions.
Since the HERA data included in PDF fits extends to $x$ as small as $x\sim2\cdot10^{-5}$,
it is obvious to expect that fixed-order perturbation theory is not sufficient to describe those data accurately.

In such a case, the resummation of small-$x$ logarithms becomes mandatory.
The theory for small-$x$ resummation in the context of collinear QCD factorization has been developed long ago
by various groups~\cite{Salam:1998tj,Ciafaloni:1999yw,Ciafaloni:2003kd,Ciafaloni:2003rd,Ciafaloni:2007gf,Ball:1995vc,Ball:1997vf,Altarelli:2001ji,Altarelli:2003hk,Altarelli:2005ni,Altarelli:2008aj,Thorne:1999sg,Thorne:1999rb,Thorne:2001nr,White:2006yh},
and it has recently been revived in Refs.~\cite{Bonvini:2016wki,Bonvini:2017ogt,Bonvini:2018xvt,Bonvini:2018iwt},
where a number of improvements have been introduced, and most importantly a public code, \texttt{HELL}~\cite{hell}, has been released.
As a representative result, we show in Fig.~\ref{fig:Pgg} (right) the $P_{gg}$ and $P_{qg}$ splitting functions
at NNLO+NLL, compared to their fixed-order counterpart.
Especially for $P_{gg}$, we see that the resummed result is very different from the fixed NNLO result
for $x\lesssim 10^{-3}$. 
Therefore, we expect to see strong deviations between resummed PDFs and NNLO ones at small $x$.

\section{\boldmath PDF fits with small-$x$ resummation}

\begin{figure}[t]
  \centering
    \includegraphics[width=0.45\textwidth]{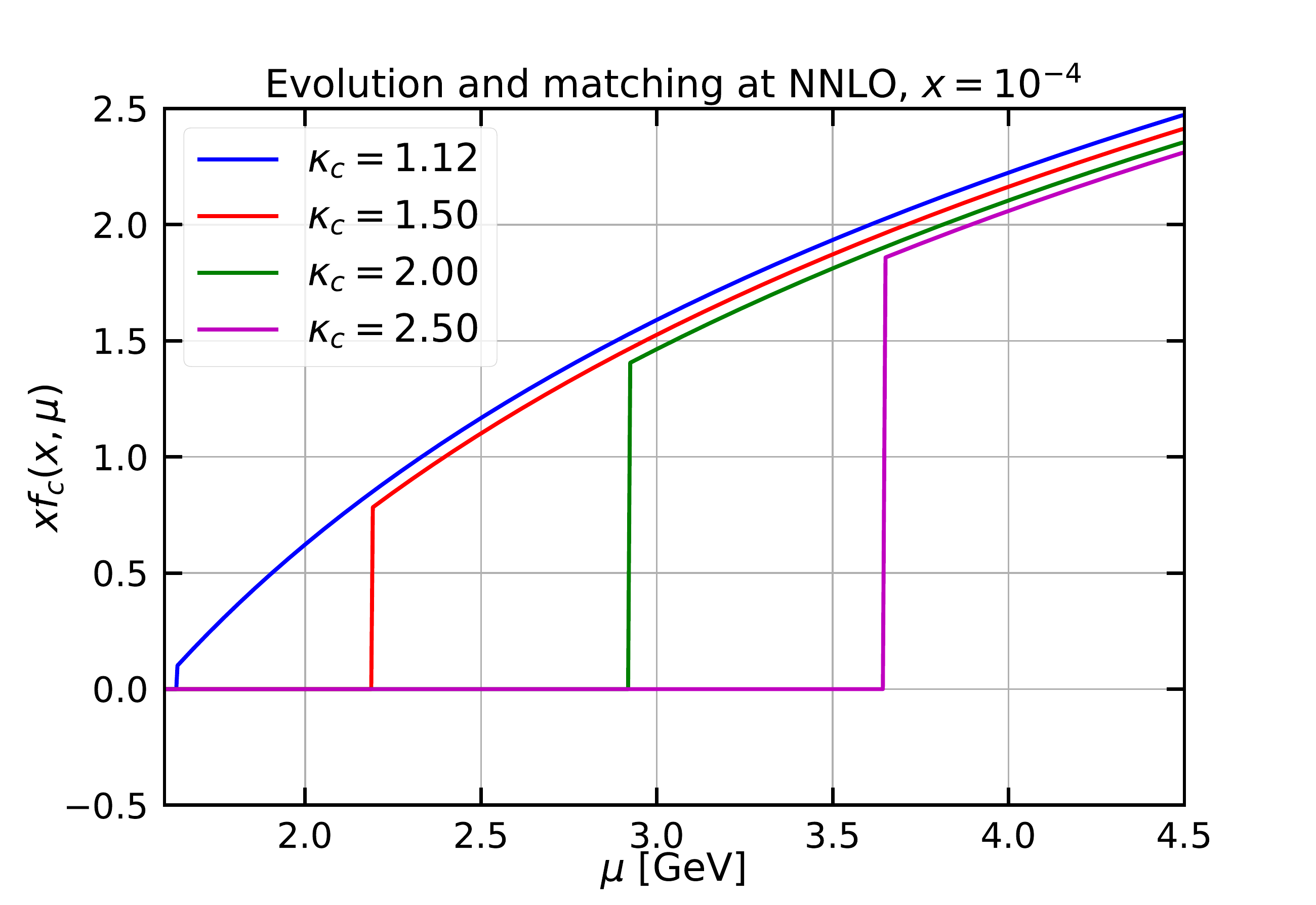}
    \includegraphics[width=0.45\textwidth]{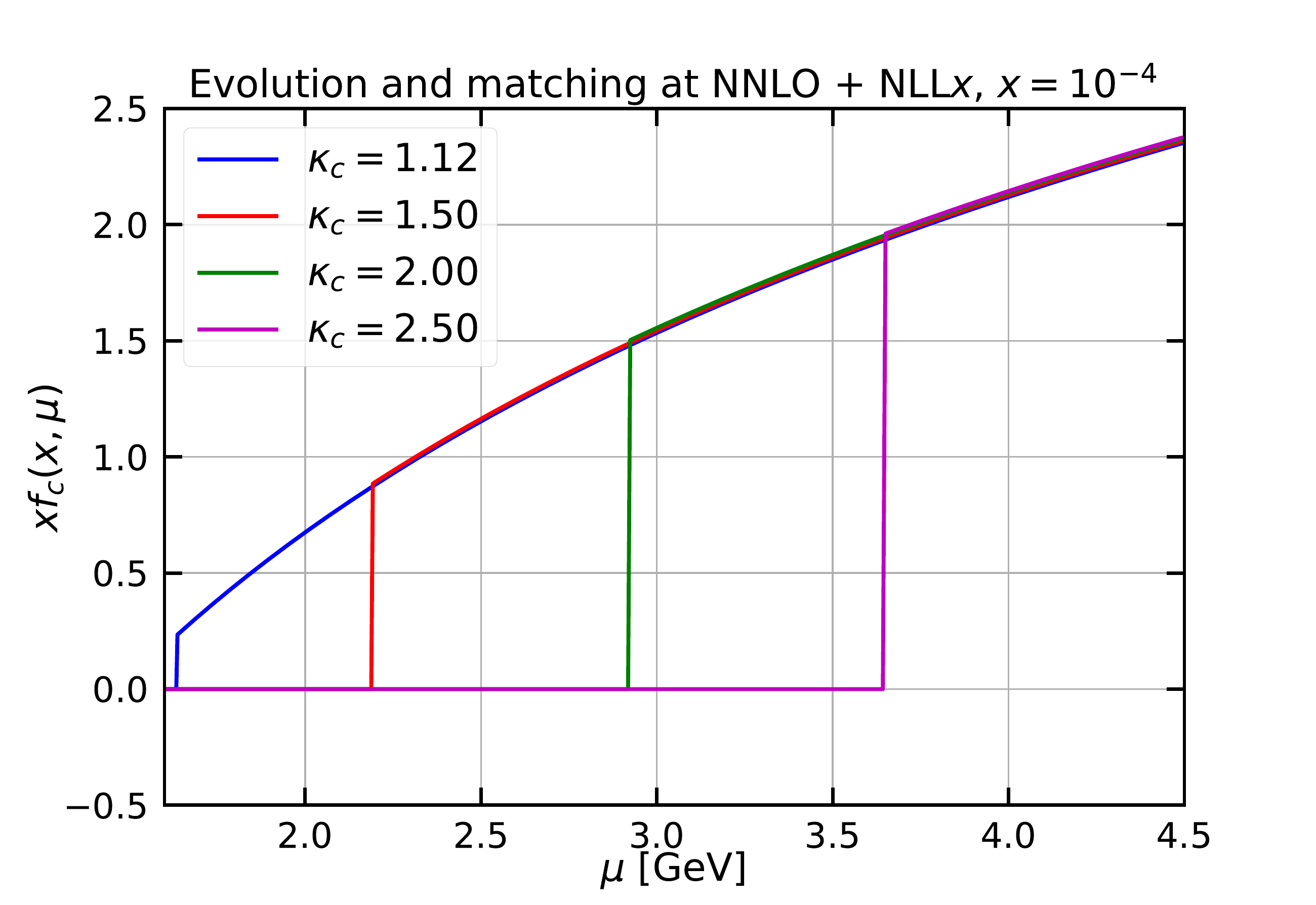}
    \caption{The charm PDF at $x=10^{-4}$ as a function of the factorization scale $\mu$
      for different values of the charm matching scale $\mu_c = \kappa_c m_c$,
      at fixed NNLO (left) and resummed NNLO+NLL (right).}
    \label{fig:matching}
\end{figure}

The public code \texttt{HELL} made the inclusion of small-$x$ resummation in a PDF fit particularly straightforward,
and indeed two studies on the inclusion of small-$x$ resummation in PDF fits have been performed~\cite{Ball:2017otu,Abdolmaleki:2018jln}
(see also Ref.~\cite{White:2006yh} for an older fit including small-$x$ resummation).
These two analyses, performed in the context of the NNPDF methodology and with the xFitter public tool, respectively,
differ in several aspects as summarized in Tab.~\ref{tab:comparison}.
The NNPDF3.1sx fit is more complete as it includes a large variety of non-HERA data, and it has been performed
both at NLO and NNLO (with and without resummation).
The xFitter study is instead focussed on the HERA data, without potential contamination from other experimental data,
and on the NNLO, where the effect of resummation is larger.
Among the various differences, the most important from the theoretical point of view is the treatment of the charm
PDF, which is fitted in the former and perturbatively generated in the latter.
This is known to lead to substantially different charm PDFs at fixed order~\cite{Ball:2016neh,Ball:2017nwa}.
One of the most remarkable outcomes of the comparison of these fits is that the extracted charm PDF is very similar
once small-$x$ resummation is included in the fit, thereby suggesting that the difference at fixed-order
was due to higher order corrections at small-$x$, not included in the perturbative-charm case
and ``fitted'' in the fitted-charm case.
Indeed, the perturbatively generated charm PDF is for instance much less dependent on the (unphysical) charm matching scale
when resummation is included than at fixed order, see Fig.~\ref{fig:matching}.

\begin{table}[h]
  \centering
  \begin{tabular}[t]{p{0.5\textwidth}p{0.45\textwidth}}
    NNPDF3.1sx \cite{Ball:2017otu} & xFitter analysis \cite{Abdolmaleki:2018jln} \\
    \midrule
    NeuralNet parametrization of PDFs & polynomial paramterization \\
    MonteCarlo uncertainty & Hessian uncertainty \\
    VFNS: FONLL & VFNS: FONLL\\
    charm PDF is fitted & charm PDF perturbatively generated \\
    DIS+tevatron+LHC ($\sim 4000$ datapoints) & only HERA data ($\sim1200$ datapoints) \\
    NLO, NLO+NLLx, NNLO, NNLO+NLLx & NNLO, NNLO+NLLx
  \end{tabular}
  \caption{Comparison of the NNPDF-like and HERAPDF like fit settings}
  \label{tab:comparison}
\end{table}

\begin{figure}[t]
  \centering
  \includegraphics[width=0.395\textwidth,page=1]{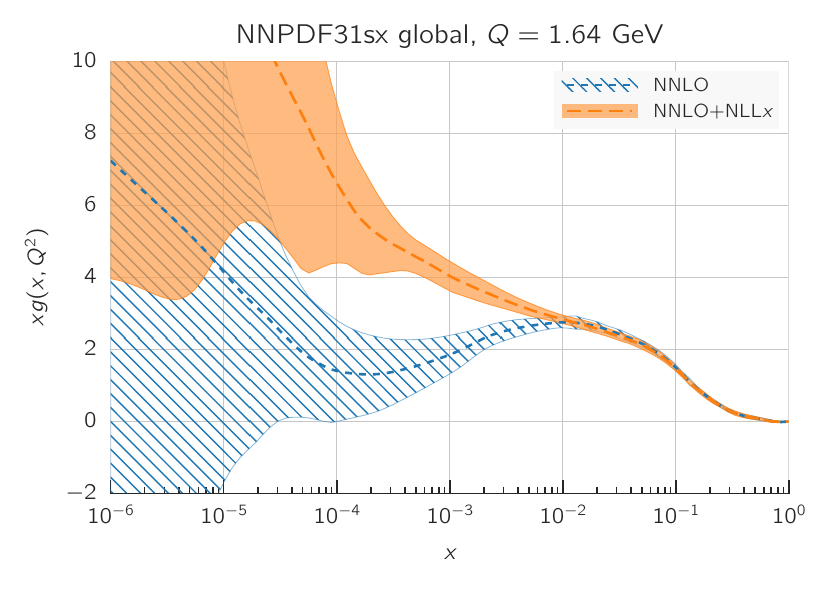}
  \includegraphics[width=0.27\textwidth,page=1]{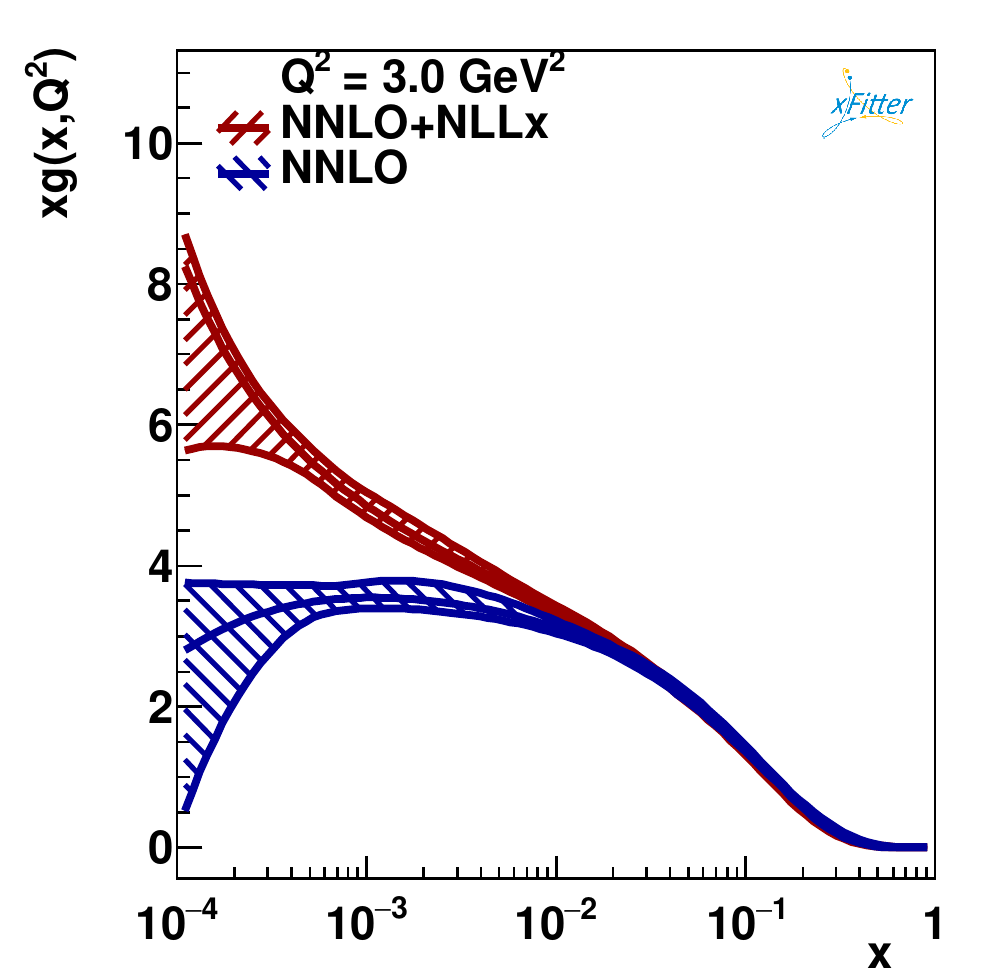}
  \includegraphics[width=0.32\textwidth,page=1]{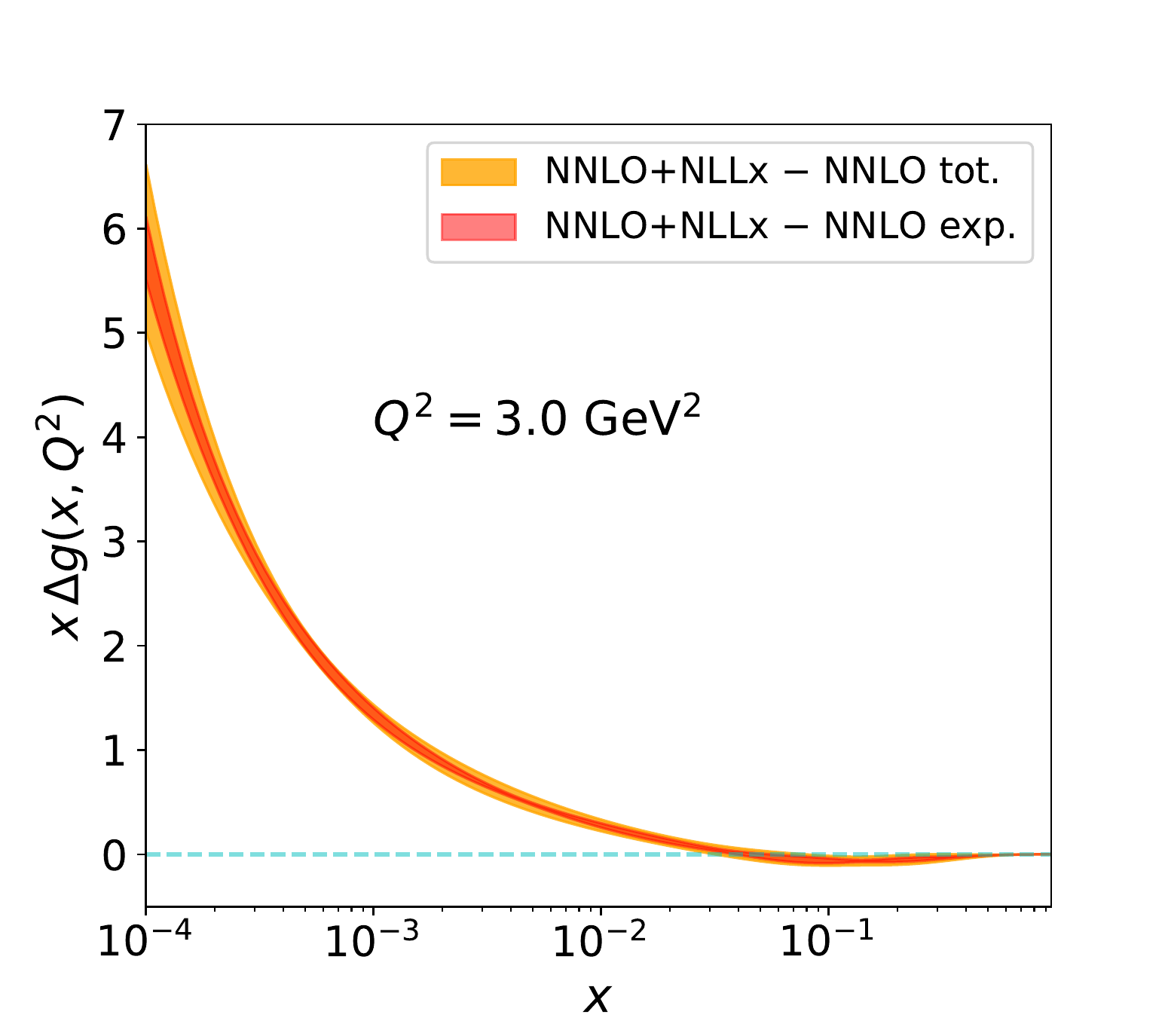}
  \caption{The gluon PDF at NNLO and NNLO+NLL from Ref.~\cite{Ball:2017otu} (left) and Ref.~\cite{Abdolmaleki:2018jln} (middle),
    and the difference between the resummed and fixed-order gluon from Ref.~\cite{Abdolmaleki:2018jln} (right).}
  \label{fig:gluon}
\end{figure}

The result of the fits are in general in good agreement.
Both studies find a significant reduction of the $\chi^2$ when resummation is added on top of NNLO,
while the quality of the fit does not change when adding resummation to NLO, as expected from the considerations
on the behaviour of splitting and coefficient functions.
The description of the HERA data at small $x$ improves,
and in particular the famous turnover~\cite{Abramowicz:2015mha} of the inclusive data at small $x$ is well reproduced.
Non-singlet PDFs are mostly unaffected by the inclusion of resummation, while the quark-singlet and most importantly
the gluon PDFs do change quite dramatically.
In Fig.~\ref{fig:gluon} we see a comparison of the NNLO and NNLO+NLL gluon PDF at a small scale,
from the NNPDF3.1sx study (left) and the xFitter study (middle).
In addition, to better quantify the significance of the impact of resummation,
the difference between the resummed and fixed-order gluon (xFitter) PDF is shown in the right plot of the figure,
which shows that the effect is very significant and large.
The most apparent difference in the gluon PDF without and with resummation is that the NNLO gluon
is rather low at small $x$, becoming also negative (within the uncertainty) for $x\lesssim10^{-4}$,
while the resummed gluon is much harder, rising steeply at small $x$ and never becoming negative (in the region constrained by data).

\begin{figure}[t]
  \centering
  \includegraphics[width=0.45\textwidth,page=1]{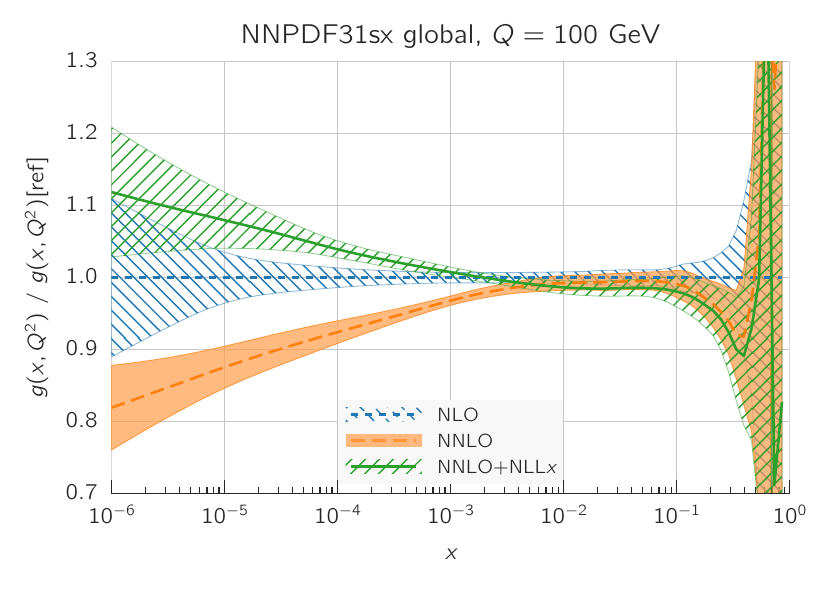}
  \hspace{\stretch{1}}
  \includegraphics[width=0.42\textwidth,page=1]{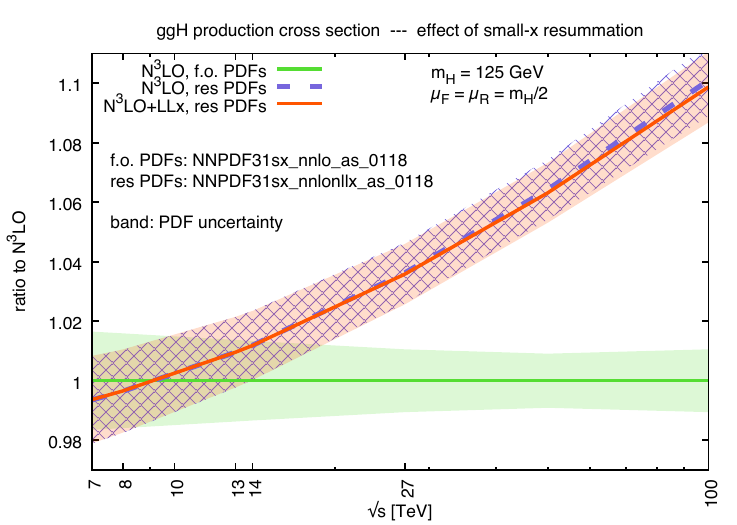}
  \caption{Left: the gluon PDF at NLO, NNLO and NNLO+NLL from Ref.~\cite{Ball:2017otu} at $Q=100$~GeV.
    Right: impact of small-$x$ resummation on the gluon-fusion Higgs cross section as a function of the collider energy.}
  \label{fig:gluon100}
\end{figure}

Such a difference survives also at higher scales, as shown in Fig.~\ref{fig:gluon100} (left) at $Q=100$~GeV.
Here, also the NLO gluon is shown. As clear from the plot, the NNLO+NLL gluon is not very different from the NLO one,
while the difference with the NNLO gluon is much larger and significant, as expected from the splitting function behaviour.
The origin of such a difference
is thus not due to the resummed gluon being strongly enhanced by the resummation,
but to the NNLO gluon being strongly depleted by the large unresummed logarithms.
The resummation restores a perturbative behaviour, but when comparing resummed results
with fixed-order ones contaminated by large unresummed logarithms the effect appears to be dramatic.
It is then clear that resummed theory allows for a more reliable description of the small-$x$ region,
and also that fixed-order theory may lead to unreliable results.
This observation has important consequences for phenomenology.

\section{Applications}

Given the significant difference between the NNLO PDFs (commonly used for most LHC phenomenological applications)
and their resummed counterparts, it is of utmost importance to understand the impact of small-$x$ resummation for
LHC observables.
As a first step in this direction, the effect of small-$x$ resummation for Higgs production in gluon fusion has been investigated
in Refs.~\cite{Bonvini:2018ixe,Bonvini:2018iwt}.
From Fig.~\ref{fig:gluon100} (right) we see that resummed theory predicts a larger cross section than the NNLO one.
The effect is small, approximately 1\% at LHC, but becomes larger as the collider energy is increased,
as smaller values of $x$ become accessible, reaching about 4\% at future HE-LHC, and up to 10\% at FCC.
Most of the effect comes from the use of resummed PDFs in place of the NNLO ones.
This is a consequence of this observable being an inclusive cross section;
we expect the effect of resummation in the coefficient functions to be more marked in differential distributions.
The inclusion of differential observables in \texttt{HELL}, starting from Drell-Yan rapidity and transverse momentum distributions,
is work in progress.

Another important application concerns the prediction of ultra high-energy (UHE) neutrino-nucleus cross sections in the atmosphere,
which can be sensitive to very small $x$, down to approximately $10^{-8}$.
In the current PDF determinations the PDF uncertainty is very large there, but they can be constrained by the inclusion of
some LHCb data, see e.g.\ Ref.~\cite{Gauld:2015yia,Gauld:2016kpd}.
This study shows that the inclusion of resummation in the prediction of the UHE neutrino-nucleus cross section
can lead up to a 15\% effect~\cite{Bertone:2018dse}.


\bibliography{references.bib}

\end{document}